\documentclass[aps,prl,twocolumn,showpacs,groupedaddress]{revtex4}
\usepackage{graphicx}

\begin{document}


\title{Feshbach-resonance-induced atomic filamentation and quantum pair correlation
in atom-laser-beam propagation}
\author{Weiping Zhang, Chris P. Search, Han Pu, Pierre Meystre
and Ewan M. Wright} \affiliation{Optical Sciences Center,
University of Arizona, Tucson, AZ 85721}

\date{\today}

\begin{abstract}
We study the propagation of an atom laser beam through a spatial
region with a magnetic field tuned to a Feshbach resonance. Tuning
the magnetic field below the resonance produces an effective
focusing Kerr medium that causes a modulational instability of the
atomic beam. Under appropriate circumstances, this results in beam
breakup and filamentation seeded by quasi-particle fluctuations,
and in the generation of correlated atomic pairs.
\end{abstract}

 \maketitle

It has been recognized for quite some time that the two-body
interactions that govern the dynamics of atomic Bose-Einstein
condensates (BECs), and which are expressed most simply in terms
of the $s$-wave scattering length $a$, play a role for matter
waves analogous to that of a cubic (Kerr) nonlinearity in optics.
This was illustrated in a number of experiments that include
matter-wave four-wave mixing \cite{phillips}, the generation of
dark \cite{dark} and bright \cite{bright} atomic solitons,
coherent matter-wave amplification \cite{amplify}, and the
generation of quantum-correlated atomic pairs \cite{ketterle2}. It
has also been demonstrated that the strength of this nonlinearity
can be varied, and its sign changed, using in particular Feshbach
resonances \cite{feshbach1,feshbach2}. In the language of
nonlinear optics, changing the sign of the scattering length
amounts to switching from a defocusing (for repulsive
interactions, $a > 0$) to a focusing (for attractive interactions,
$a < 0$) nonlinear medium. The tunability of the scattering length
offers a degree of flexibility of considerable interest in the
study of the manybody properties of atomic condensates,  as well
as for nonlinear atom optics applications.

As pointed out by Timmermans {\em et al.} \cite{timmermans},
caution is in order when considering the dynamics of a condensate
in the vicinity of a Feshbach resonance. In the atomic systems
under consideration, they correspond to a situation where the
total energy of two colliding atoms equals the energy of a bound
molecular state. In the vicinity of the resonance, it is essential
to properly account for the molecular dynamics. Indeed, it has
recently been unambiguously established experimentally that a
proper description of the system requires that one accounts for
the coherent coupling between the atomic and molecular field,
resulting in the onset of atom-molecule coherence in BECs
\cite{atommolecule}. The presence of the molecular state is also
predicted to lead to ``resonant superfluidity'' in
quantum-degenerate atomic Fermi systems \cite{resonanceSF}.

Far from resonance, however, it is possible to adiabatically
eliminate the molecular field from the system's dynamics, in which
case the atomic condensate is subject to an effective, tunable
scattering length that describes the binary atomic collisions. The
goal of this letter is to illustrate how this tunable scattering
length can be exploited to manipulate and control the propagation
of a continuous-wave atom laser beam. The specific geometry that
we consider consists of an atom laser beam propagating through a
region where a spatially dependent magnetic field is applied to
switch the nature of the binary collisions from being repulsive,
or defocusing, to attractive, or focusing. This results in a
spatial change of the interatomic interaction similar to the
presence of a Kerr medium in nonlinear optics, albeit with some
important differences. We show that under appropriate
circumstances, the quasi-particle fluctuations in the initial atom
laser beam can seed the filamentation of the atom laser beam, with
the generation of pair-correlated atoms in spatial sidebands. This
is similar to the amplification of squeezed light by parametric
processes in a nonlinear optical Kerr medium, but with the
important difference that in the present situation it is not
necessary to start from a squeezed input.

We recall that the generation of correlated atomic pairs from BECs
has recently been studied by a number of authors
\cite{pu,ketterle2}. In these works, correlated atomic pairs are
generated either from vacuum fluctuations \cite{pu}, or grow via
four-wave mixing from the combined effects of a seed beam and
vacuum fluctuations \cite{ketterle2}. In contrast, the proposed
scheme involves the generation of a pair-correlated atomic source
from \textit{quasi-particle vacuum fluctuations}, with an atom
laser beam as the sole pump beam.

A Feshbach resonance is a scattering resonance where two particles
in an open channel collide to form a bound state in the closed
channel. In the case of quantum-degenerate atomic fields, such as
Bose-Einstein condensates or atom laser beams, the bound state in
the closed channel is a coherent molecular state. The dynamical
equations describing the atom-molecule coherent coupling can be
written as \cite{timmermans}
\begin{eqnarray}
i\hbar \frac{\partial \hat{\Phi}_a}{\partial t} &=&
\left(-\frac{\hbar^2 \nabla^2}{2m} + U_a \hat{\Phi}_a^{\dagger}
\hat{\Phi}_a + U_{am} \hat{\Phi}_m^{\dagger} \hat{\Phi}_m \right)
\hat{\Phi}_a \nonumber \\
&&\;\;+ 2 \alpha \hat{\Phi}_m \hat{\Phi}_a^{\dagger}
\nonumber \\
i\hbar \frac{\partial \hat{\Phi}_m}{\partial t} &=&
\left(-\frac{\hbar^2 \nabla^2}{4m} + \epsilon  + U_m
\hat{\Phi}_m^{\dagger} \hat{\Phi}_m + U_{am}
\hat{\Phi}_a^{\dagger} \hat{\Phi}_a \right) \hat{\Phi}_m \nonumber
\\ &&\;\;+ \alpha \hat{\Phi}_a^2.
\end{eqnarray}
Here $\hat{\Phi}_{a,m}$ are the atomic and molecular field
operators, $m$ is the atomic mass, $U_{a}= (4\pi \hbar^2/m)a_a$,
$U_m=(2\pi \hbar^2/m)a_m$, and $U_{am}=(3\pi \hbar^2/2m)a_{am}$
are the atom-atom, molecule-molecule, and atom-molecule binary
interaction strengths, with $a_a$, $a_m$, and $a_{am}$ the
corresponding scattering lengths, and $\alpha$ is the coupling
strength the Feshbach resonance. The parameter $\epsilon$ is the
energy difference between the molecular state and the atomic
state.

For large resonance detunings $|\epsilon|$, it is possible to
adiabatically eliminate the molecular field, leading to the
effective Heisenberg equation of motion
\begin{equation}
i\hbar \frac{\partial \hat{\Phi}_a}{\partial t} =
\left(-\frac{\hbar^2 \nabla^2}{2m} + \hbar \chi
\hat{\Phi}_a^{\dagger} \hat{\Phi}_a \right) \hat{\Phi}_a,
\label{atomnonlinear}
\end{equation}
for the atomic Schr{\"o}dinger field. In this expression, the
effective coupling constant $\hbar \chi \equiv U_a -2
\alpha^2/\epsilon$ accounts for the change of the the strength of
the atomic binary collisions resulting from the Feshbach
resonance.

We now proceed to study the propagation of an atom laser through a
spatially-dependent effective nonlinearity, which for simplicity
we take to be of the form
$$\hbar \chi(z) \equiv U_a -2 (\alpha^2/\epsilon
)\Theta(z),$$
where $\Theta(z)$ is a Heaviside step function. In particular, we
consider the situation in which the binary collisions change from
repulsive for $z<0$, to attractive for $z>0$. The continuous wave
atom laser beam is incident from $z<0$, and we describe the
incident atomic field as a coherent beam of Bose-condensed atoms
propagating along the $z$-axis with wave vector $k_0 {\hat{\bf z}}
= (mv_0/\hbar) {\hat {\bf z}}$. We also ignore any atoms that are
back reflected at the interface: If we take account of the fact
that the magnetic field actually varies over a length $L$ around
$z=0$, then using a WKB analysis we find that back reflections are
negligible if $(2\pi/k_0)(2\alpha^2/|\epsilon|)n_0/L \ll v_0^2/2m$
where $n_0$ is the density of atoms in the condensate.

An important distinction between this and the generic situation
considered in nonlinear optics, where the beam is incident from
vacuum, is that the incident atomic field is self-interacting.
Hence, a proper description must include the quasi-particle vacuum
fluctuations propagating together with the coherent BEC component.
As such, a realistic atom laser beam consists of the sum of a
moving condensed component, described classically, and quantum
fluctuations \cite{fetter},
\begin{equation}
\hat{\Phi}_a({\bf r},t) = \big[ \sqrt{n_0} + \hat{\psi}({\bf r},t)
\big] e^{i(k_0-k_n) z -i\omega_0 t} , \qquad z<0
,\label{atomlaser1}
\end{equation}
where $\sqrt{n_0}$ describes the coherent BEC component (taken
real without loss of generality), $\hbar
\omega_0=\hbar^2k_0^2/2m$, and $k_n=\chi n_0/v_0$ is a nonlinear
phase shift. The noise operator $\hat{\psi}({\bf r},t)$ obeys
bosonic commutation relations. From the corresponding Heisenberg
equation of motion (\ref{atomnonlinear}), one observes that the
binary collisions contribute a term of the form
$$i\hbar\frac{\partial\hat{\psi}({\bf r},t)}{\partial t} \sim U_a n_0
\left (\hat{\psi}({\bf r},t)+ \hat{\psi}^\dagger({\bf r},t)\right
)$$
to the evolution of the noise operator $\hat{\psi}({\bf
r},t)$. This form of evolution is quite familiar in quantum
optics, where it is known to lead to the generation of squeezed
states. Thus, the repulsive self-interaction of the incident atom
laser beam for $z<0$ is sufficient to provide initial squeezed
state vacuum fluctuations at the input port of the (focusing)
nonlinear medium $z=0$.

We proceed by expanding the noise operator $\hat{\psi}({\bf r},t)$
in terms of plane-waves as
\begin{eqnarray}
\hat{\psi}({\bf r},t) = \int d{\bf q} \int_0^{\infty} d{k_z} \big[
\hat{a}_0({\bf q},k_z,t) e^{ik_z
(z-v_0t)} \nonumber\\
-\hat{b}_0({\bf q},k_z,t) e^{-ik_z (z-v_0t)} \big] e^{i {\bf q}
\cdot {\bf r}_{\perp}} ,\label{atomlaser2}
\end{eqnarray}
where ${\bf q}$ is the transverse wave vector and the operators
$\hat{a}_0({\bf q},k_z,t)$ and $\hat{b}_0({\bf
q},k_z,t)=-\hat{a}_0({\bf q},-k_z,t)$ describe forward and
backward propagating modes in the rest frame of the atomic beam,
respectively.

In the incident region with repulsive interactions, $z<0$, it is
convenient to express the small quantum fluctuations about the
condensate wave function in terms of quasi-particles characterized
by bosonic annihilation operators $\hat{\alpha}({\bf q},k_z,t)$
and $\hat{\beta}({\bf q},k_z,t)=-\hat{\alpha}(-{\bf q},-k_z,t)$,
given by \cite{fetter}
\begin{eqnarray*}
\hat{a}_0({\bf q},k_z,t) &\equiv & u_k \hat{\alpha}({\bf q},k_z,t)
+ v_k \hat{\beta}^{\dagger}({\bf q},k_z,t) ,\\ \hat{b}_0({\bf
q},k_z,t) & \equiv & u_k \hat{\beta}(-{\bf q},k_z,t) + v_k
\hat{\alpha}^{\dagger}(-{\bf q},k_z,t),
\end{eqnarray*}
where
\begin{eqnarray*}
u_k^2 = v_k^2+1=\frac{1}{2}\left( \frac{\epsilon_k+\hbar\chi
n_0}{E_k}+1 \right) ,
\end{eqnarray*}
with $k \equiv \sqrt{q^2+k_z^2}$, $\epsilon_k \equiv \hbar^2
k^2/2m$, and $E_k=\sqrt{\epsilon_k^2 +2\epsilon_k \hbar\chi n_0}$
being the quasiparticle energy associated with $\hat{\alpha}({\bf
q},k_z,t)$ and $ \hat{\beta}({\bf q},k_z,t)$.

We are now in a position to study the propagation of the atom
laser beam as it enters the region of Feshbach resonance, $z >0$.
Specifically, in this paper we focus on the initiation of the
filamentation process before appreciable depletion. For this
purpose, it is sufficient to consider a linearized perturbation
theory in the slowly varying amplitude approximation. For $z>0$ we
decompose the atomic field operator in the same manner as Eq.
(\ref{atomlaser1}) with the substitution of $\hat{\Phi}({\bf
r},t)$ in place of $\hat{\psi}({\bf r},t)$ for the fluctuation
operator. Furthermore, analogous to Eq.~(\ref{atomlaser2}), we
also decompose the noise operator $\hat{\Phi}({\bf r},t)$ in terms
of quasi-plane waves,
\begin{equation}
\hat{\Phi}({\bf r},t) = \int d{\bf q}\,\hat{\Phi}_{\bf q}(z,t)\,
e^{i {\bf q} \cdot {\bf r}_{\perp}}, \label{ansatz}
\end{equation}
where
\begin{eqnarray}
\hat{\Phi}_{\bf q}(z,t)= \int_0^{\infty} &d{k_z} & \left[
\hat{a}({\bf q},k_z,z)  e^{ik_z(z-v_0 t)} \right. \nonumber \\
&&\left.+ \hat{b}({\bf q},k_z,z) e^{-ik_z (z-v_0t)} \right]
.\label{qqq}
\end{eqnarray}
Here we allow for a slow $z$-dependence of the operators $\hat{a}$
and $\hat{b}$. This approach, familiar from quantum optics, allows
one to account for the possible amplification of the quantum
noise. Spatial variation of the quasi-particle operators was
ignored in the region $z<0$ due to the absence of sideband gain
for repulsive binary collisions.

Substituting Eq. (\ref{ansatz}) into Eq.~(\ref{atomnonlinear})
yields pairs of coupled operator equations that describe the
linearized spatial evolution of the condensate sidebands,
\begin{widetext}
\begin{equation}
i \frac{\partial}{\partial T} \left[
\begin{array}{c}
\hat{a}({\bf q},k_z,z)\\
\hat{b}^{\dagger}(-{\bf q},k_z,z)
\end{array} \right] =
\left[ \begin{array}{cc} \big ( \epsilon_q
+\chi n_0 \big ) & \chi n_0 \\
-\chi n_0 & - \big (\epsilon_q + \chi n_0 \big )
\end{array} \right]
\left[ \begin{array}{c}
\hat{a}({\bf q},k_z,z)\\
\hat{b}^{\dagger}(-{\bf q},k_z,z)
\end{array} \right].
\label{sideband}
\end{equation}
where $T=t+z/v_0$. Assuming a solution to Eq.~(\ref{sideband}) of
the form $e^{gT}$ and matching the field operator solutions in the
two regions at $z=0$ we obtain finally
\begin{eqnarray}
&&\hat{a}({\bf q},k_z,z) = \Big \{ \big[U(q,z)u_k + V( q,z)v_k
\big] \hat{\alpha}({\bf q},k_z) +  \left [ U(q,z)v_k + V(q,z)u_k
\right ]\hat{\beta}^{\dagger}({\bf q},k_z) \Big \},
\nonumber\\
&&\hat{b}({\bf q},k_z,z) = -\Big \{ \big[ V(q,z)u_k + U( q,z)v_k
\big] \hat{\alpha}^{\dagger}(-{\bf q},k_z) + \big[ V(q,z)v_k +
U(q,z)u_k \big] \hat{\beta}(-{\bf q},k_z) \Big \}
,\label{solutions}
\end{eqnarray}
\end{widetext}
where the coefficients $U$ and $V$ are given by
\begin{eqnarray}
U(q,z)&=& \cosh \left(g \frac{z}{v_0} \right) - \frac{i}{g}
\left(\epsilon_q + \chi n_0 \right) \sinh \left(g \frac{z}{v_0}\right), \nonumber\\
V(q,z)&=& i \frac{\chi n_0}{g} \sinh\left (g \frac{z}{v_0}\right
), \label{UV}
\end{eqnarray}
with the gain parameter
\begin{equation}
g = \sqrt{(\chi n_0)^2 -\left (\epsilon_q+ \chi n_0\right )^2}.
\label{gain}
\end{equation}
Without loss of generality we have taken $t=0$ in order to focus
on how the filamentation develops with propagation distance.

We observe that the linearized solution of Eqs.~(\ref{solutions})
can exhibit gain in the case of a focusing Feshbach resonance,
$\chi < 0$. Specifically, the gain coefficient $g$ is real for $q
\le 2 \sqrt{m|\chi|n_0/\hbar}$, with a maximum, $g_{\rm max}=
|\chi|n_0$, for the transverse wavevector
\begin{equation} q_{\rm max} =
\sqrt{2m|\chi|n_0/\hbar}. \label{qmax}
\end{equation}
Physically, this growth of a range of coupled pairs of spatial
sidebands $\pm{\bf q}$ around the condensate corresponds to a
transverse modulational instability with a cone of unstable
transverse momenta peaked around $|{\bf q}|=q_{\rm max}$. If
allowed to grow this instability will ultimately lead to the
breakup of the incident laser beam into transverse structures or
filaments similarly to the situation in nonlinear optics
\cite{campillo,boyd}. If instead of using a plane-wave condensate
we consider a beam with widths $w_{x,y}$ in the $x$ and $y$
transverse directions, we can eliminate the instability in the
$y$-direction, for example, by choosing $q_{\rm max}w_y<1, q_{\rm
max}w_x \gg 1$, so that the most unstable wavevector cannot be
supported in the $y$-direction. In that case the cone of unstable
wavevectors becomes a pair of sidebands that are peaked at $\pm
q_{\rm max}$ along the $x$-direction. This quasi-one-dimensional
geometry may be better to isolate correlated atomic pairs.

This modulational instability is intimately related to the
collapse of $^{85}$Rb condensates seen in recent experiments
\cite{feshbach2}. In these experiments, stable repulsive BEC's
with densities of $n_0\approx 10^{12}$ cm$^{-3}$ are produced in
the presence of a magnetic field tuned close to a Feshbach
resonance ($B \approx 160$ G) and then the magnetic field is tuned
above $167$ G where the interaction becomes attractive. Far from
the resonance the interactions are attractive with a background
scattering length of $a_{bg}=-23.85$ nm \cite{atommolecule}. Using
these parameters gives a $q_{\rm max}=7.75\times 10^{6}$ m$^{-1}$
or a transverse velocity of $0.6$ cm/s for the sideband atoms.

Proceeding with the example of a plane-wave condensate, to
quantify the statistical properties of the generated atomic
sidebands, we evaluate the field correlation and intensity
correlation functions for ${\hat \Phi}_{\bf q}(z)$ defined in
Eq.~(\ref{qqq}). Assuming that the quantum fluctuations in the
incident atom laser beam are in the quasi-particle vacuum state
$|\Psi_0\rangle$, i.e., $\hat{\alpha} |\Psi_0 \rangle= \hat{\beta}
|\Psi_0 \rangle =0$, and with Eqs.~(\ref{solutions}) and
(\ref{UV}) we find,
\begin{eqnarray}
\langle \hat{\Phi}^{\dagger}_{\bf q} \hat{\Phi}_{{\bf q}'}
\rangle &=& \rho(q) \,\delta_{{\bf q},{\bf q}'} ,\label{normal} \\
\langle \hat{\Phi}_{\bf q}^{\dagger}\hat{\Phi}_{\bf
q}\hat{\Phi}_{\bf q'}^{\dagger}\hat{\Phi}_{{\bf q}'} \rangle &=&
\rho(q)\rho(q') +|Q(q)|^2\delta_{{\bf q},-{\bf q}'},
\label{anomalous}
\end{eqnarray}
where all the operators are evaluated at the same longitudinal
distance $z$ and in Eq.~(\ref{anomalous}), we have taken ${\bf q}
\neq {\bf q}'$. Physically, $\rho(q)$ is the probability density
in momentum space of finding an atom with a transverse momentum
$\hbar{\bf q}$ after propagating a distance $z$ in the Feshbach
region, and $|Q(q)|^2$ is the joint probability that if one atom
is detected with momentum $\hbar{\bf q}$ a second one will be
detected with opposite momentum $-\hbar{\bf q}$. They are given
by,
\begin{widetext}
\begin{eqnarray}
\rho(q) &=& 2 \int_0^{\infty} dk_z \left [ v_k^2 + \frac{\chi^2
n_0^2 }{g^2} (u_k^2+v_k^2) \sinh^2\left (g \frac{z}{v_0}\right )
-\frac{2\chi n_0}{g^2} \varepsilon_q  u_k v_k
\sinh^2\left (g \frac{z}{v_0}\right ) \right ], \nonumber \\
 Q(q) &=& -2 \int_0^{\infty} dk_z \left \{ u_k v_k + \left [
\frac{\chi n_0}{g^2} (u_k^2+v_k^2) - \frac{2}{g^2}\varepsilon_q
u_k v_k \right ]  \left [ \varepsilon_q \sinh^2\left (g
\frac{z}{v_0}\right )+ \frac{ig}{2} \sinh\left (2 g
\frac{z}{v_0}\right ) \right ] \right \}, \label{qspectra}
\end{eqnarray}
\end{widetext}
with $\varepsilon_q \equiv \epsilon_q + \chi n_0$.

\begin{figure}
\includegraphics*[width=0.85\columnwidth,
height=1.0\columnwidth]{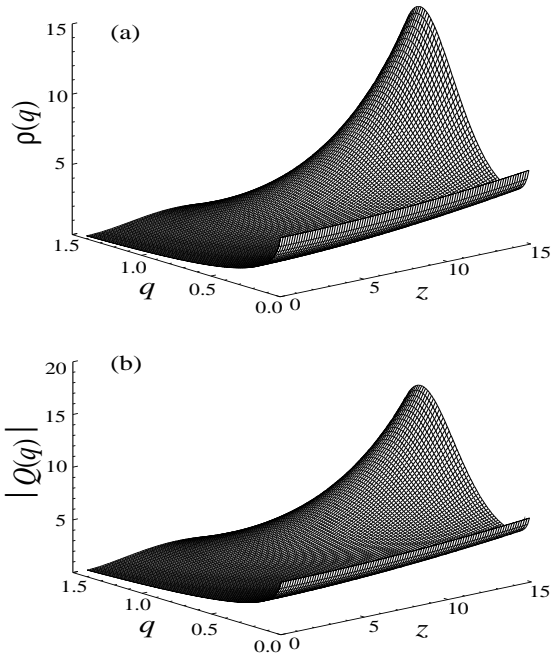} \caption{Correlation spectrum of
the correlated atomic pairs calculated from Eq.~(\ref{qspectra}).
Due to the divergence at large wave number, the upper limit of
$k_z$ is cut off at $k_s=mc_p/\hbar$---the wave number
corresponding to the Bogoliubov sound velocity. Here the wave
numbers, energy and length are in units of $k_s$, $\hbar^2
k_s^2/(2m)$ and $1/k_s$, respectively. The parameters used in the
calculation are $v_0=5c_p$, $\chi n_0=-0.5$. The plots show that
both $\rho$ and $|Q|$ grow fastest near $q_{\rm max}=1/\sqrt{2}$
[see Eq.~(\ref{qmax})].} \label{fig3}
\end{figure}

As an illustration we assume that atoms are detected after having
propagated a distance $z$ in the Feshbach region such that the
linearized analysis is still valid. The rate at which atoms are
detected at an angle $\theta=\tan^{-1}(q/k_0)$ relative to the
$z$-axis will be proportional to $\rho(q)$ while the rate of
coincidence detections at $+\theta$ and $-\theta$ will be
proportional to $|Q(q)|^2$. The corresponding probability
densities are plotted in Fig.~\ref{fig3},
from which we see that, as expected physically, with increasing
propagation distance $z$ the most probable transverse wavevector
for detecting either an off-axis atom or a correlated atomic pair
at $\pm {\bf q}$ is increasingly $|{\bf q}|=q_{\rm max}$. These
results show that the transverse modulational instability can
indeed be an effective source of correlated atomic pairs.

In conclusion, we have studied the propagation of an atom laser
beam through a Feshbach resonance region where the effective
interaction between atoms is tuned from being repulsive to being
attractive. The attractive interaction is equivalent to a
self-focusing Kerr medium for the atom laser beam which can
generate quantum-pair-correlated atomic filament beams by a
modulational instability. This opens up new window to create
nonclassical ultracold atomic sources by coherent control of the
atomic interactions. The quantum-pair-correlated atomic beam may
have potential applications in both fundamental research and
applications in quantum information and computation.

This work is supported in part by the US Office of Naval Research
under Contract No. 14-91-J1205 and No. N00014-99-1-0806, by the
National Science Foundation under Grant No. PHY98-01099, by the US
Army Research Office, and by the Joint Services Optics Program.

\end{document}